\documentclass[twocolumn,aps,prl,floats,superscriptaddress,showpacs]{revtex4}
\usepackage{amsmath}
\usepackage{here}
\usepackage{epsfig}
\usepackage{graphicx}
\usepackage{dcolumn}
\usepackage{bm}

\def\gapp{\lower.35em\hbox{$\stackrel{\textstyle>}{\sim}$}}
\def\lapp{\lower.35em\hbox{$\stackrel{\textstyle<}{\sim}$}}

\begin{document}
\bibliographystyle{apsrev}

 \title{Valley polarized electronic beam splitting in graphene}

\author{J.L. Garcia-Pomar$^{1}$, A. Cortijo$^{1}$, and M. Nieto-Vesperinas}

\address{Instituto de Ciencia de Materiales de Madrid, Consejo Superior de Investigaciones
Cientificas, Campus de Cantoblanco, Madrid 28049, Spain.}
\email{jlgarcia@icmm.cisc.es}
\email{cortijo@icmm.csic.es}
\email{mnieto@icmm.csic.es}

\begin{abstract}
We show how the trigonal warping effect in doped graphene can be
used to produce fully valley polarized currents. We propose a
device that acts both as a beam splitter and a collimator of these
electronic  currents. The result is demonstrated trough an optical
analogy using two dimensional photonic crystals.
\end{abstract}
\pacs{73.23.Ad, 85.75.-d, 42.70.Qs, 42.79.Fm, 81.05.Uw, 73.63.-b}
 \maketitle
Since its recent synthesis \cite{Netal05} graphene is proving to
be a fruitful system in which ideas coming from varied areas of
physics successfully merge: soft condensed matter
\cite{fasolino07}, \textit{pseudorelativistic} quantum
electrodynamics \cite{katsnelson06} and, more recently, optics
\cite{cheiakov06} and \textit{valleytronics}
\cite{valley_beenakker07}. The concept of \textit{valleytronics}
\cite{valley_beenakker07} describes the use of the valley
degeneracy in graphene as a degree of freedom to carrying
information in the same way as the spin does in spintronic. In the
scheme proposed in \cite{valley_beenakker07} the generation and
detection of the valley polarized current are achieved by means of
a quantum point contact in a graphene ribbon.
 The main difficulty in the experimental
realization of this approach comes from the necessity of tailoring
the graphene samples into nanoribbons and quantum point contacts
with zigzag edges needed to support edge states. In this letter we
propose an alternative mechanism for getting a different response
from each valley using  the effect of the trigonal warping
\cite{dress00} that can be tuned by a gate voltage in graphene
samples of arbitrary shapes.
%
\begin{figure}
  \begin{center}
    \epsfig{file=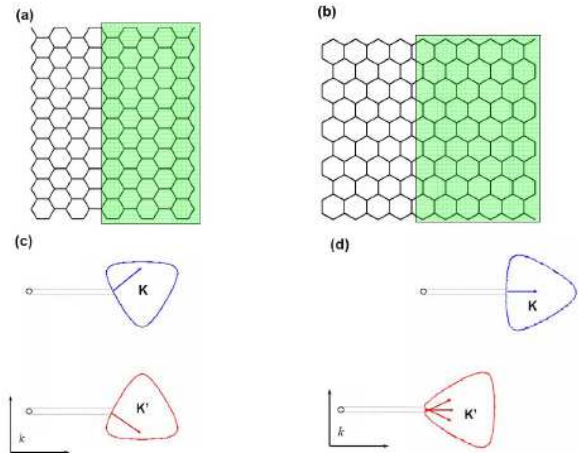,height=6cm}
    \caption{(Color online) The two limiting cases for the deposition of the
    gate over the graphene sample. The green area denotes
    the region doped with holes at energies where the TW becomes relevant. (a) Interface of the gate
    parallel to the zigzag orientation and (b) the armchair interface. (c) and (d) represent the corresponding
    isofrequency lines showing the conservation of the parallel wavevector for the zigzag and armchair interface,
    respectively. We obtain two directions of refraction for the zigzag orientation parallel
    to the interface depending on whether the chosen valley is $K$ or $K'$. For the armchair
    interface, a forward direction of refraction appears due to
     the little dispersion produced by the second valley.}
    \label{composicio}
\end{center}
\end{figure}
The main idea relies on the fact that at energies where the
trigonal warping is noticeable, the Fermi surfaces around each
valley are different and anisotropic. This allows us to suggest a
device  based on a p-n junction where the response is valley
asymmetric. (Figs. 1 and 2).


In the tight-binding approximation of the low energy band
structure for graphene, the dispersion relation including both the
Dirac and the TW terms, is:

\begin{equation}
E_{s}(\mathbf{k})=\pm\hbar
v_{F}\sqrt{k^2+\frac{a^2}{16}k^{4}+\frac{a}{2}s\left(k_{x}^{3}-3k_{x}k_{y}^2\right)},\label{dispersion}
\end{equation}
where $v_{F}$ is the Fermi velocity, $a$ is the lattice constant
and the short notation $k^2=k_{x}^2+k_{y}^2$ is used. The
parameter $s=\pm1$ refers to the two valleys.

The average current moving trough the sample is
$\langle\mathbf{J}_s\rangle=\frac{e}{\hbar}|t_s|^2\mathbf{v}_{g,s}$,where
$t_s$ represents the transmission coefficient and
$\mathbf{v}_{g,s}$ stands for the group velocity in each valley.
When the TW term is considered, $\mathbf{v}_{g,s}$ has the form
\begin{equation}
\mathbf{v}_{g,s}=\frac{1}{\hbar}\nabla_{\mathbf{k}}E_{s}(\mathbf{k})=\pm\frac{v^2_F}{2E_{s}(\mathbf{k})}\left((2+\frac{a^2}{4}k^2)\mathbf{k}+\frac{a}{2}s\mathbf{F}\right),\label{groupvelocity}
\end{equation}
where $\mathbf{F}=(3k_x^2-3k_y^2,-6k_x k_y)$. When TW is
considered, $|t_s|^2$ does not essentially differ from that
obtained when only the linear term is considered in the whole
$p-n$ junction \cite{katsnelson06,Cheianov_prb06} having almost
the same shape for both valleys. The major role in the different
behavior of $\langle\mathbf{J}_s\rangle$ for each valley is played
by the group velocity $\mathbf{v}_{g,s}$, since it explicitly
depends on $s$.

%
\begin{figure}
  \begin{center}
    \epsfig{file=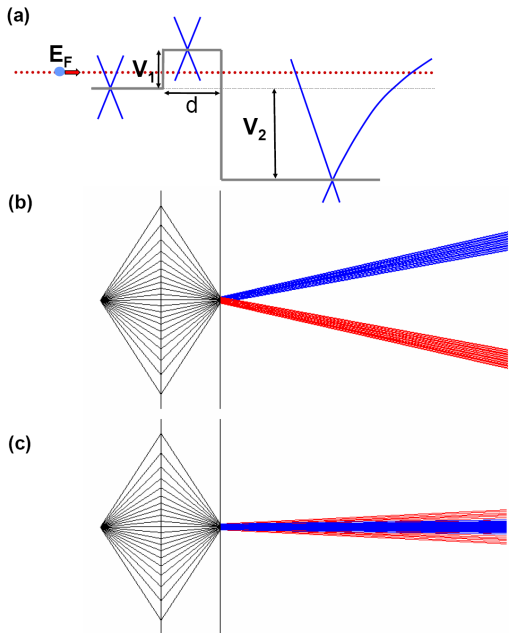,width=7cm}
    \caption{(Color online) (a) Potential profile of a $n-p-n^{-}$ junction
    and schematic band structure of one valley.
    (b) Theoretical ray tracing of the beams coming from a point source in a
    junction with the interface parallel to the zigzag orientation.
    $E_{F}=0.05eV$, $V_{1}=0.10eV$ and $V_{2}=-1.5eV$. The beam
    splitting is noticeable. (c) Armchair junction. We now obtain a collimation perpendicular
    to the interface.}
    \label{rayoszig}
\end{center}
\end{figure}

It is well known that optical and electronic systems share many
wave-like properties. A $p-n$ junction in graphene has an photonic
analogy with an optical system composed of two effective media
with opposite signs of their refractive index. In optical material
slabs with negative refractive index, Snell's law leads to the
phenomenon of light focussing \cite{Veselago68,Pendry00}. A
similar behavior has been proposed in graphene for electrons
\cite{cheiakov06}. If the Fermi level is chosen such that the
conduction band, (for which the group velocity $\mathbf{v}_{g}$
points \textit{outwards} the energy surface), in one side of the
junction, is connected to the valence band in the other side,
(where $\mathbf{v}_{g}$ now points \textit{inwards} the energy
surface), the system behaves as the electrical counterpart of a
Veselago Lens \cite{Veselago68}.

In Fig.\ref{composicio} we show a field effect $n-p^+$ device in
graphene. In the $p^+$ area, the applied gate voltage is such that
the chemical potential is located at energies where the TW effect
becomes relevant (according to ARPES data \cite{Lanzara_nat06}
this appens for energies of $0.6-0.7 eV$). When the TW term is
considered, the band is no longer isotropic around each K point.
We have then to care about the relative orientation between the
lattice and the deposited gate. Due to the symmetry of the
honeycomb lattice, all the possible orientations lie between the
two limiting cases shown in Figs.\ref{composicio}(a),(b).
Fig.\ref{composicio}(c) shows the direction of $\mathbf{v}_{g,s}$
determined by conservation of the momentum parallel to the
potential barrier interfaces when the boundary is parallel to the
zigzag orientation. Due to the $s$ dependence of
$\mathbf{v}_{g,s}$ the currents associated to each valley travel
in different directions leading to two different currents
polarized in the valley index. Fig.\ref{composicio}(d) represents
the parallel momentum conservation when the gate is oriented along
the armchair direction. The current associated to one of the
valleys (valley $K$) is fully collimated because, by momentum
conservation, the accessible area of the energy surface is almost
plane. The current belonging to the other valley is more dispersed
because the accessible area has a higher  curvature.
\begin{figure}
  \begin{center}
    \epsfig{file=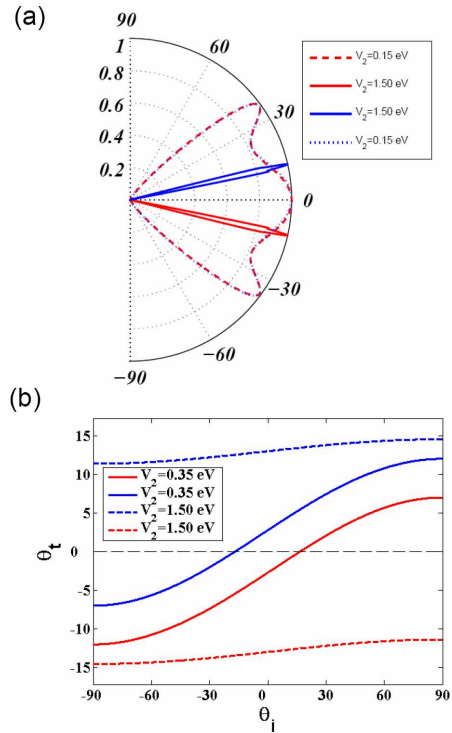,width=7cm}
    \caption{(Color online)
    (a)Transmission probability for the K (blue lines) and K'(red lines)  as a function of the output
    angle $\theta_{t}$ for different negative potentials $V_{2}$.
    The parameters are:  barrier  width:  $d=100nm$,
    initial energy:   $E_{F}=0.05eV$, initial
    potential: $V_{1}=0.14eV$.
    (b) Output angle $\theta_{t}$  of the
    electronic beam  as a function of the incident angle
    $\theta_{i}$ for valleys  K (upper blue lines) and K'(lower red
    lines). (See text).
    Barrier width: $100nm$, initial
    energy $E_{F}=0.05eV$, first potential step: $V_{1}=0.10eV$.}
    \label{beamfotonic}
\end{center}
\end{figure}
%
With these characteristics, we can design a device to achieve the
realization of a beam splitter with an $n-p-n^{-}$ junction (Fig.
2(a)) with the interface parallel to the zigzag orientation. In
the first zone, we introduce an electron source. In this region,
the Fermi level $E_{F}$ is at low energy in the conduction band,
where the Fermi surface is circular and the linear Dirac regime is
valid. This area is separated from a zone of low chemical
potential ($p$-region) in the valence band, by a barrier with a
potential $V_{1}$  . In the interface we have both negative
refraction and focusing \cite{cheiakov06}. The last step with
negative potential $V_{2}$ is placed immediately before the
electron focusing in order to get a very small illuminated zone.
In this way, we have narrower exit beams avoiding the formation of
caustics which would appear if the position of the focus were
reached before the last interface. Finally, the last region is at
high voltages in the conduction band ($n^{-}$-region) i.e. it has
a strong TW. A polarization of valley occurs, this giving rise to
a beam splitting of the electronic beams belonging to different
valleys (Fig.2(b),(c)). Fig.3 shows the two beams obtained for a
potential $V_{2}=-1.5 eV$ (TW case) $25.4^{\circ}$ apart from each
other for maximum of transmission. The beam dispersion is
approximately only $3^{\circ}$. In the Dirac regime (lower
energies) we observe that the splitting of the beams is reduced.

\begin{figure}
  \begin{center}
    \epsfig{file=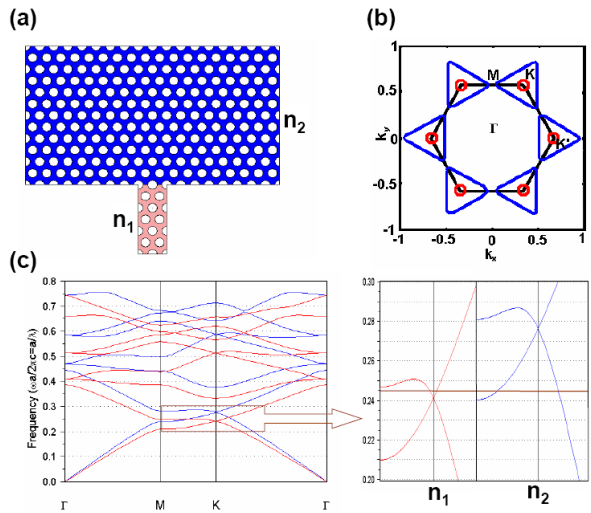,width=9.25cm}
    \caption{(Color online) (a) Geometry of the trigonal photonic crystal (PC) of air holes and
    optical analogy of a p-n junction in graphene. (b) Brillouin zone and isofrequency lines for the two
    PCs for $n_{1}$ (red circles) and for $n_{2}$ (blue triangles).
    (c) Band diagram for a trigonal photonic crystal of air holes with
    radius $r=0.33a_{PC}$ in a dielectric matrix with refractive index $n_{1}=3.04$
  (red bands) and
    $n_{2}=2.65$ (blue bands). The two insets show details of the zone around the $K$
    point where we observe that for the wavelength $\lambda=4.08a_{PC}$,  $n_{1}$ corresponds to the conduction
    band, whereas $n_{2}$ is located at the TW valence band.}
    \label{fotonico}
\end{center}
\end{figure}
\begin{figure}
  \begin{center}
    \epsfig{file=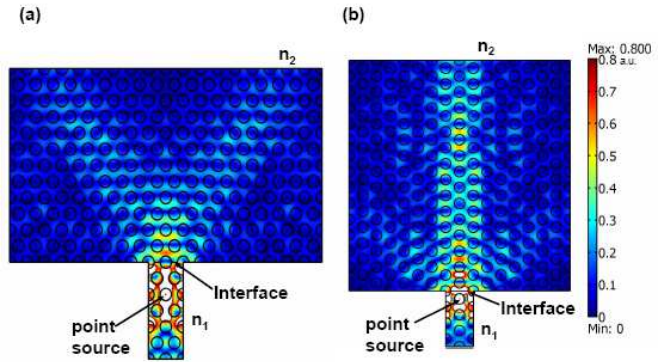,width=8.75cm}
    \caption{(Color online) Maps of the  electric field modulus of
    a wave originated at a point source located inside a photonic
    crystal with refraction index $n_{1}$
    and propagating through  another photonic crystal of dielectric matrix
    $n_{2}$.
    (a) Armchair orientation, illustrating the beam splitting of the two
    different valleys K and K'. (b) Zigzag orientation. We can
    see a collimation of the beams since the curvature in the trigonal distorted band is
    smaller than the one of the circular isoline in the proximity of the Dirac point.
    The angular dispersion is low.}
    \label{fotonico2}
\end{center}
\end{figure}
A different situation occurs if the interface is parallel to the
armchair orientation. In this configuration,(Fig. 1(b)) the beam
of one valley is strongly collimated and waveguided (blue color in
Fig.2(b)) due to the flat portion of the isofrecuency line. The
other valley slightly disperses the beam. Due to the big
difference of curvatures between the Dirac and the TW isoenergies,
the allowed wavevectors $\textbf{k}$ are situated in a zone of the
TW isoline with small curvature involving a low dispersion (red
color in Fig.2(b)).

Up to day, the most efficient way of identifying graphene samples
is by optical contrast \cite{Novoselov04}. The maximum of this
contrast depends both on the thickness of the substrate and on the
wavelength of the monochromatic radiation used to illuminate the
sample, usually in the visible range. For this range, the maximum
contrast corresponds to a thickness of $300nm$ and the values
reached for the chemical potential are of the order of $0.3 eV$.
However,  it is theoretically found \cite{Castrooptics07} that
another maximum in the contrast appears at a thickness of
$90-100nm$. With this range, it is possible to get values of the
chemical potential up to $0.6 eV$, just in the threshold where the
TW starts to be appreciable. The effect discussed here can then be
effectively observed with minor experimental changes in current
samples, like increasing the gate voltage and decreasing the
substrate thickness.

{\it  An optical analogy.}
Control of beam propagation, collimation and focusing in photonic
crystals (PCs) is already well developed
\cite{Kosaka,Notomi,Wang,Garcia-Pomar,Rakich06}, thus one can
assess the validity of the above analysis for graphene by putting
forward its analogy with this photonic system. In fact,an optical
analogy of the graphene system was previously described
\cite{Haldane05,beenakker07}. We can realize it with a 2-D PC with
honeycomb, kagome or trigonal structure where the gap between the
first and second bands is replaced by a Dirac point in the $K$
point of symmetry. As an example, we have used a trigonal lattice,
with parameter $a_{PC}$, of air cylinders of radius $r=0.33a_{PC}$
inside a dielectric matrix with refractive index $n$
(Fig.\ref{fotonico}(a)). We notice that, in the case of the
trigonal lattice, the Brillouin zone (BZ) (Fig.\ref{fotonico}(b))
coincides with the real lattice, while in the graphene honeycomb
lattice the BZ is rotated 30 degrees with respect to the real
lattice, which implies that the zigzag and armchair situations
described in graphene are exchanged in the PC. Now we cannot
change the light frequency throughout the crystal like we change
the chemical potential in a transistor of field effect in
graphene, but we can displace the frequency position of the Dirac
point in the band diagram by slightly varying the refractive index
of the dielectric matrix (Fig.\ref{fotonico}(c)). In this way, a
junction of two PCs with the same trigonal structure but with
slightly different refractive index, mimics the field effect
transistor in graphene.

Fig.\ref{fotonico2} shows a finite elements numerical simulation
of the electric field modulus distribution from a point source,
emitting TE polarized light (i.e. the electric field is parallel
to the long axis of the air holes) with a wavelength
$\lambda=4.08a_{PC}$ in a dielectric matrix with refractive index
$n_{1}=3.04$ together with another PC of the same structure but a
slightly smaller refractive index $n_{2}=2.65$ for the two
limiting cases: armchair(Fig.\ref{fotonico2}(a)) and zigzag
(Fig.\ref{fotonico2}(b)). The creation of a beam splitter in the
latter PC for the armchair situation is illustrated by the fact
that the two isofrequency curves in the K and K' points are
triangles pointing in opposite directions from each other (right
and left). Conversely, if we put the zigzag structure parallel to
the interface, we obtain a collimation of the beams.

To summarize, we have proposed an efficient way to get
\textit{valley polarized} beams using the trigonal warping effect
in the graphene bands. We have shown a possible design of a beam
splitter and a collimator device. This formation of valley
polarized beams, is a key ingredient for the use of this new
polarization in graphene as an effective information carrying
degree of freedom. Besides, the transport properties of graphene
can constitute a useful and novel device in future
nanoelectronics. Finally, we have checked the validity of these
phenomena, by means of an optical analogy with two dimensional
photonic crystals and proposed a theory of isofrequency lines as a
useful tool to simulate these kind of processes in graphene
systems.

\section*{Acknowledgements} We gratefully acknowledge illuminating
discussions and useful comments of Prof. Mar\'{i}a A. H.
Vozmediano and Ramon Aguado. Work supported by the Spanish DGICYT
and the European Union. J.L. G.-P. acknowledges I3P grant program.
A. C. acknowledges MEC of Spain for financial support.





%

\bibliography{graphenix_ref}

\end{document}